\documentclass[10pt,conference]{IEEEtran}
\usepackage{graphicx}
\usepackage{textcomp}
\usepackage{fancyhdr}
\usepackage[hidelinks]{hyperref}
\usepackage{braket}
\usepackage[T1]{fontenc}
\usepackage{algorithm}
\usepackage{algpseudocode} 
\usepackage{tikz}
\usepackage{caption}

\usepackage{cite}
\usepackage{amsmath,amssymb,amsfonts}
\usepackage{physics}
\usepackage{threeparttable}
\usepackage{booktabs}
\usepackage{xcolor}
\usepackage{url}
\title{ZAP: Zoned Architecture and Performant Compiler for Field Programmable Atom Array}

\author{
Chen Huang\textsuperscript{$\dagger$,1,2,6}, Xi Zhao\textsuperscript{$\dagger$,1,2,7}, Hongze Xu\textsuperscript{$\dagger$,1,2}, Weifeng Zhuang\textsuperscript{1,2}, \\
Meng-Jun Hu\textsuperscript{1,2,a}, Dong E. Liu\textsuperscript{1,2,3,4,5,b}, Jingbo Wang\textsuperscript{1,2,c}%
\thanks{Corresponding authors: Mengjun Hu (humj@baqis.ac.cn), Jingbo Wang (wangjb@baqis.ac.cn), Dong E. Liu (dongeliu@mail.tsinghua.edu.cn).}
\\
\IEEEauthorblockA{\textsuperscript{1}Beijing Academy of Quantum Information Sciences, Beijing 100193, China\\
\textsuperscript{2}Beijing Key Laboratory of Fault-Tolerant Quantum Computing, Beijing 100193, China\\
\textsuperscript{3}State Key Laboratory of Low Dimensional Quantum Physics,\\
Department of Physics, Tsinghua University, Beijing 100084, China\\
\textsuperscript{4}Frontier Science Center for Quantum Information, Beijing 100184, China\\
\textsuperscript{5}Hefei National Laboratory, Hefei 230088, China\\
\textsuperscript{6}Department of Computer Science and Engineering, The Chinese University of Hong Kong, Hong Kong SAR, China\\
\textsuperscript{7}CAS Key Laboratory of Quantum Information, University of Science and Technology of China, Hefei 230026, China\\
\textsuperscript{$\dagger$}These authors contributed equally to this work.\\
Email: \textsuperscript{a}humj@baqis.ac.cn, \textsuperscript{b}dongeliu@mail.tsinghua.edu.cn,
\textsuperscript{c}wangjb@baqis.ac.cn}
}

\begin{document}
\maketitle
\pagestyle{plain}

\begin{abstract}
    The scalability of neutral-atom quantum computing is increasingly limited by a compiler--architecture challenge: logical circuits must be mapped onto dynamically reconfigurable atom arrays while controlling crosstalk, transport overhead, and hardware constraints. To address this problem, we present ZAP, a co-designed zoned architecture and deterministic compiler for field-programmable atom arrays. ZAP partitions the array into storage and entanglement zones and combines hardware-aware ASAP-separate scheduling, look-ahead placement, and conflict-aware routing in a single-pass compilation flow, thereby avoiding the repeated global search used in prior approaches. Evaluated on structured quantum benchmarks and random 3-regular circuits, ZAP consistently delivers multi-order-of-magnitude compilation speedups while maintaining competitive or superior execution quality. Relative to ZAC and PowerMove, ZAP typically reduces compilation time from tens of seconds to below 0.1~s and achieves speedups exceeding 1,000$\times$; relative to Enola, the speedup exceeds 10,000$\times$ on the evaluated suite. ZAP's fidelity gains are most pronounced on structured workloads with irregular connectivity and nonuniform qubit reuse, where its scheduling and placement decisions more effectively suppress crosstalk and limit transport-related loss, while on random circuits it remains competitive and preserves the same scalability advantage. These results show that hardware-structured, non-iterative compilation provides a practical path toward fast, scalable, and noise-aware neutral-atom quantum computing.
\end{abstract}

\section{Introduction}

In recent years, quantum computing has seen significant advancements in both software and hardware~\cite{preskill2018quantum,montanaro2016quantum,bruzewicz2019trapped,harrow2009quantum,kjaergaard2020superconducting}. These improvements have expanded the computational capabilities of quantum systems and opened up potential applications across various fields~\cite{lekitsch2017blueprint,mcardle2020quantum,cao2019quantum,aumasson2017impact,fernandez2020towards,choi2019tutorial}. Driven by the advances in laser technology, neutral atom arrays have recently emerged as a leading platform for scalable quantum computing~\cite{henriet2020quantum, wintersperger2023neutral,cong2022hardware,graham2022multi,bluvstein2024logical}, demonstrating systems with thousands of physical qubits~\cite{manetsch2024tweezer} and high-fidelity gate operations~\cite{evered2023high}. Unlike architectures with static qubits, such as superconducting circuits, neutral atom systems feature dynamic qubit connectivity: atoms can be physically moved and repositioned in real-time using optical tweezers~\cite{bluvstein2022quantum}. This architectural capability eliminates the need for costly SWAP gate chains but introduces a formidable hardware-software co-design challenge~\cite{murali2019full}: how to efficiently compile a quantum circuit to orchestrate thousands of mobile atoms.
  
The core of this challenge lies in taming the immense search space of potential atom placements and movements. Existing compilers for this paradigm, such as Enola~\cite{tan2024compilation} and ZAC~\cite{lin2025reuse}, attack this problem with slow, iterative, search-based algorithms, while PowerMove~\cite{ruan2024powermove} is non-iterative in placement but remains iterative in routing. While functional for small circuits, these approaches fundamentally fail to scale. Compilation times explode from minutes to hours for modestly sized problems, creating an untenable bottleneck for future fault-tolerant systems and severing the rapid feedback loop essential for quantum algorithm research. This compilation overhead, coupled with physical constraints like atom transport time and crosstalk, forms a critical barrier to unlocking the full potential of neutral atom quantum computers. 

To dismantle this barrier, we introduce ZAP, a performant compiler framework tailored for zoned neutral atom architectures. ZAP's central insight is to impose structure on the hardware to simplify the compilation task. We spatially partition the atom array into two distinct regions: a compact storage zone for idle qubits and an optimized entanglement zone where all two-qubit gates are executed. This physical separation of concerns is the key to our performance, significantly decreasing crosstalk. ZAP employs a deterministic, non-iterative placement algorithm that positions active qubits in the entanglement zone with minimal movement. This is combined with a hardware-aware scheduling policy: within a standard ASAP framework, ZAP separates the scheduling of two-qubit and single-qubit gates to reduce unnecessary movement in zoned architectures. Coupled with look-ahead placement and routing, this policy balances parallelism against crosstalk and atom transport overhead. The result is a system where the hardware architecture and compiler work in concert, transforming a computationally intractable mapping problem into a deterministic and highly efficient process.

Our approach yields transformative improvements in both compilation speed and execution quality. By eliminating iterative placement, ZAP achieves compilation speedups of multiple orders of magnitude compared with prior compilers. On a 100-qubit circuit, compilation completes in under 0.1 seconds on a laptop, whereas other compilers require minutes or even hours. These speedups are accompanied by fidelity gains driven by reduced atom movement and mitigated crosstalk.
  
Our main contributions are summarized as follows:
\begin{itemize}
    \item \textbf{Fast and deterministic placement for zoned architectures:} We develop a deterministic, single-pass placement algorithm tailored to the topology of zoned architectures. Beyond minimizing movement distance, the placement heuristic incorporates a lightweight routing-aware preference that favors movement patterns with fewer anticipated AOD conflicts, thereby removing the critical iterative-search bottleneck found in prior work.

    \item \textbf{Hardware-aware scheduling \& routing for zoned architectures:} Within a standard ASAP framework, we adopt a separate scheduling policy for two-qubit and single-qubit gates and couple it with look-ahead placement and routing decisions. Unlike prior zoned compilers that either always retain qubits in the entanglement zone or always return them after gate execution, ZAP introduces a hybrid idle-qubit management strategy that dynamically balances crosstalk exposure against transport overhead.

    \item \textbf{Strong empirical performance:} We demonstrate that ZAP's co-design reduces compilation times from minutes to seconds or less while simultaneously improving execution quality.
\end{itemize}

This paper is organized as follows. Sec.~\ref{sec:background} reviews neutral atom quantum computing. Sec.~\ref{sec:hardware} presents our proposed zoned architecture and the compilation challenges it addresses. Sec.~\ref{sec:fidelity} introduces our fidelity analysis. Sec.~\ref{sec:zap} describes the ZAP compiler framework, including its scheduling, placement, and routing algorithms. Sec.~\ref{sec:setup} and Sec.~\ref{sec:results} present the experimental setup and evaluation results. Sec.~\ref{sec:related-work} discusses related work, and Sec.~\ref{sec:conclusion} concludes the paper.

\section{Background}\label{sec:background}
  
\subsection{Quantum Computing}
  
Quantum computing operates by manipulating quantum bits (qubits), the fundamental units of quantum information. Unlike a classical bit, a qubit can exist in a superposition of two basis states, $\ket{0}$ and $\ket{1}$, expressed as
\begin{equation}
    \ket{\psi} = \alpha \ket{0} + \beta \ket{1},
\end{equation}
where $\alpha$ and $\beta$ are complex probability amplitudes satisfying the normalization condition $|\alpha|^2+|\beta|^2=1$. 
  
Quantum algorithms are composed of a series of quantum gate operations. Single-qubit gates, such as the Hadamard gate, are used to create superposition, while two-qubit gates, such as the controlled-NOT (CNOT) gate, are used to establish entanglement between qubits. According to the Solovay-Kitaev theorem~\cite{dawson2005solovay}, any complex quantum operation can be approximated to arbitrary precision using a finite set of single- and two-qubit gates. This ``universal gate set'' forms the foundation for constructing all quantum algorithms.
  
\subsection{Neutral Atom Quantum Platform}
  
In neutral atom quantum computing, qubits are typically represented by two stable internal hyperfine ground states of an atom, corresponding to the logical states $\ket{0}$ and $\ket{1}$. This approach leverages quantum mechanical principles to enable precise control of qubit operations through laser pulses.
  
Single-qubit gates are achieved by inducing transitions between the hyperfine states using laser pulses, where the pulse duration and phase determine the operation parameters. These operations can be described by unitary matrices of the form
\begin{equation}
    U_{\text{single}} = 
    \begin{pmatrix}
        \cos \theta & -\mathrm{i} \mathrm{e}^{\mathrm{i} \phi} \sin \theta \\
          -\mathrm{i} \mathrm{e}^{-\mathrm{i} \phi} \sin \theta & \cos \theta
    \end{pmatrix},
\end{equation}
where $\theta$ and $\phi$ are controlled through laser modulation. High-fidelity single-qubit operations have been demonstrated on neutral atom platforms, ensuring their reliability for quantum computation.
  
A key challenge in neutral atom quantum computing is the implementation of high-fidelity two-qubit gates, which rely on the Rydberg blockade effect. When an atom is excited to a high-energy Rydberg state, it exerts a strong dipole-dipole interaction with surrounding atoms, which ``blocks'' neighboring atoms within a certain radius from being excited to the same Rydberg state. This conditional mechanism is the physical basis for realizing controlled two-qubit gates, such as the controlled-Z (CZ) gate, which can be represented by the matrix
\begin{equation}
    U_{\text{CZ}} = 
    \begin{pmatrix}
        1 & 0 & 0 & 0 \\
        0 & 1 & 0 & 0 \\
        0 & 0 & 1 & 0 \\
        0 & 0 & 0 & -1
    \end{pmatrix}.
\end{equation}
The CZ gate can be transformed into a CNOT gate by applying Hadamard gates to the target qubit before and after the CZ gate: $U_{\text{CNOT}} = (I\otimes H) \cdot U_{\text{CZ}} \cdot (I \otimes H)$, where $I$ is the identity operator on the control qubit. This transformation allows flexible implementation of two-qubit operations and is widely employed in neutral atom platforms. Therefore, high-fidelity single-qubit gates and CZ gates constitute the fundamental gate operations on neutral atom platforms~\cite{evered2023high,bluvstein2024logical}.
  
\section{Hardware Constraints}\label{sec:hardware}
  
The architectural design of a neutral atom quantum computer is fundamentally shaped by the capabilities and limitations of its underlying control hardware. This section details these physical constraints and shows how they motivate our zoned architecture, which in turn defines the core challenges for the compiler.
  
\subsection{SLM and AOD}

Qubits in neutral atom quantum computing platforms are encoded in individual atoms trapped by optical tweezers. The spatial positioning of these atoms is controlled by manipulating the optical tweezers. Two key optical devices facilitate this manipulation: the spatial light modulator (SLM) and the acousto-optic deflector (AOD). While both can trap atoms, they serve distinct and complementary roles~\cite{bluvstein2022quantum,manetsch2024tweezer}.
  
The SLM generates a fixed two-dimensional lattice of tightly focused optical traps, where each site can stably hold a single atom. These sites form the static qubit array, with deep potential wells to minimize atom loss and thermal motion during computation.
  
In contrast, the AOD provides a dynamic, reconfigurable tweezer array with adjustable spacing along orthogonal directions. Its mobility is governed by structural constraints imposed by the multi-tone frequency control mechanism: atoms can move only within their respective rows or columns, and each row or column must be displaced as a single unit within a defined movement zone. This gives rise to the equidistant, or grid, constraint: for any two atoms moved simultaneously, the projections of their trajectories onto the row and column axes must remain distinct at all times to avoid hardware conflicts. Consequently, atomic trajectories are not independent; the motion of one qubit is inherently coupled to the positions and paths of others sharing the same AOD line. This coupling turns routing into a coordinated scheduling problem and necessitates specialized protocols for collision-free parallel transport. This limited but rapid mobility is exploited to shuttle atoms between different functional zones of the processor.
  
In our architecture, as shown in Fig.~\ref{fig:tweezers}, the SLM-generated lattice provides the qubit sites for two distinct regions: a lower storage zone and an upper entanglement zone. To perform a two-qubit gate, the AODs transport selected qubits from the storage zone into the entanglement zone. Within this zone, any pair of atoms brought within the Rydberg blockade radius can execute a CZ gate via laser excitation. This ability to move qubits on demand for targeted entanglement enables universal quantum computation and is consistent with recent large-scale neutral-atom hardware demonstrations based on coherent transport and zone-based operation~\cite{bluvstein2022quantum,bluvstein2024logical,manetsch2024tweezer}.

\begin{figure}[t]
    \centering
    \includegraphics[width=0.99\columnwidth]{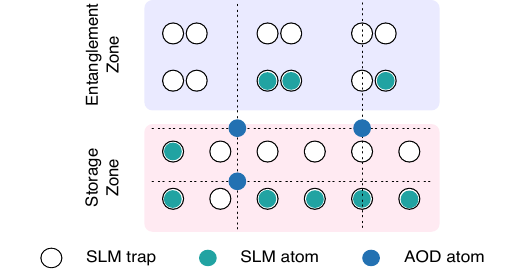}
    \caption{Zoned architecture for neutral atom quantum computing. The qubit array is divided into a storage zone and an entanglement zone. An SLM generates static optical traps (white circles) for qubit sites, while AODs create mobile tweezers (dashed lines) to transport selected atoms (blue circles) between sites. \label{fig:tweezers}}
    \label{fig:tweezers}
\end{figure}
  
\subsection{Zoned Architecture}
  
The zoned architecture is a key design paradigm in neutral atom platforms that spatially segregates quantum operations to enhance fidelity and scalability. This approach is primarily motivated by the challenges of laser control for two-qubit gates. While single-qubit gates use individually addressable lasers, implementing two-qubit gates via Rydberg blockade is more robustly achieved with a global laser field that illuminates an entire region~\cite{levine2018high, saffman2010quantum}. Such a field, however, would cause significant crosstalk by affecting idle qubits in its path. The zoned architecture solves this by partitioning the qubit array into dedicated regions, primarily a storage zone and an entanglement zone, as shown in Fig.~\ref{fig:tweezers}. This physically separates active qubits from idle ones, mitigating crosstalk and preserving fidelity.

The storage zone functions as a high-coherence, stable environment where qubits can be held idle or have single-qubit gates applied. The qubit positions are defined by a static SLM lattice with spacing chosen to reduce unwanted interactions and crosstalk between inactive qubits. This separation helps preserve coherence during circuit execution.
  
The entanglement zone is the active region where high-fidelity two-qubit gates are performed (single-qubit gates can also be applied here). This zone is precisely targeted by the Rydberg laser system needed to induce the blockade effect for CZ gates. The SLM trap sites in this region are arranged into distinct pairs. Within each pair, the two sites are separated by a distance smaller than the Rydberg blockade radius, allowing their corresponding atoms to become entangled. Conversely, the distance between different pairs is kept larger than the blockade radius to prevent unwanted interactions during concurrent gate operations. 
  
The critical link between these zones is the dynamic atom transport enabled by AODs. Qubits are shuttled from the storage zone to specific locations in the entanglement zone to perform a gate, and then moved back or rearranged for subsequent operations. This physical separation offers significant advantages. It isolates idle qubits in the storage zone from the control fields and potential stray light associated with the entanglement zone, thereby minimizing crosstalk and preserving qubit coherence. Furthermore, it facilitates a more efficient workflow, allowing for the pre-loading and rearrangement of qubits for upcoming gates while other gates are currently executing, which can improve overall parallelism and reduce circuit execution time. These capabilities are increasingly supported by recent large-scale tweezer-array and architecture-level advances~\cite{manetsch2024tweezer,bluvstein2026faulttolerant}.

\subsection{Compilation Challenges in Neutral Atom Platforms}
  
Compilation for neutral atom platforms is governed by a key physical constraint: the Rydberg blockade effect, which causes atoms to interact as qubits only when they are in close proximity. Consequently, a quantum circuit's logical qubits must be mapped to a physical arrangement of atoms that allows for these interactions. The compiler's primary objective is to orchestrate the necessary atomic movements to perform gates efficiently, minimizing execution time while improving fidelity.
  
A significant challenge in this platform is maximizing parallelism in gate execution. While single-qubit gates can be applied independently, scheduling two-qubit gates is more complex due to the constraints imposed by the Rydberg blockade~\cite{jaksch2000fast,lukin2001dipole,gaetan2009observation}. In principle, neutral atom architectures support concurrent two-qubit operations, provided that active qubit pairs are sufficiently spaced such that their blockade regions do not overlap. Maintaining independent Rydberg blockades requires that, during a two-qubit gate, no other atoms within the blockade radius are inadvertently excited to Rydberg states. This is ensured through precise laser targeting and adequate spatial separation between gate pairs. Violations of this condition can lead to undesired interactions and reduced gate fidelity. Consequently, compilation algorithms must carefully coordinate atomic placement and gate timing to maximize parallelism while preserving fidelity. This spatial constraint makes the efficient execution of large-scale circuits particularly challenging, as reducing circuit depth must be balanced against the need to prevent blockade interference.
  
A practical strategy for managing gate dependencies and optimizing parallelism in quantum circuits is to represent the circuit as a Directed Acyclic Graph (DAG), where each node corresponds to a quantum gate, and edges indicate execution constraints. This representation helps identify independent operations that can be executed in parallel, reducing the critical path length of the circuit. In neutral atom platforms, where qubit connectivity is dynamically reconfigurable, DAG-based analysis provides a structured way to plan atomic movements and optimize spatial arrangements. By mapping qubit interactions onto the physical layout, the compiler can schedule movements efficiently, minimizing unnecessary delays and mitigating coherence loss.
  
Among these challenges, this work focuses explicitly on reducing atom-movement time without compromising gate fidelity. Unlike in superconducting platforms, where some speed--fidelity trade-offs can be managed through pulse shaping or error mitigation, neutral atom systems face stricter movement-related constraints. Atomic motion introduces decoherence and heating, while prolonged interaction times increase susceptibility to noise and errors. Therefore, minimizing movement time while preserving gate fidelity is not merely an optimization goal but a fundamental requirement for maintaining overall circuit performance. Our compilation strategy employs a DAG-based scheduling approach, which we term the ASAP scheduling strategy, combined with spatial optimization. Notably, our method avoids iterative refinement in the placement process, substantially reducing compilation time. By integrating these techniques, we enhance the efficiency and reliability of neutral atom quantum circuit execution.
  
\section{Fidelity Analysis}\label{sec:fidelity}

Multiple factors influence the fidelity of quantum circuit execution on neutral atom platforms. In addition to the intrinsic fidelities of single- and two-qubit gates, the finite qubit lifetime constrains gate execution times, impacting overall fidelity. Furthermore, atom transport between SLM and AOD traps introduces heating effects and time delays, further degrading performance. However, the direct fidelity loss from the transport process itself can often be neglected. Experimental work has shown that for sufficiently controlled transport speeds, the process introduces negligible intrinsic error~\cite{bluvstein2022quantum}. Therefore, we assume the primary penalty of atom movement is the time it adds to the circuit's duration, which contributes to decoherence, rather than a direct loss of fidelity from the transport itself. Under these assumptions, the overall circuit fidelity can be estimated using the following expression:
\begin{equation}
    \begin{aligned}
        f = \left( f_1 \right)^{g_1} \cdot \left( f_2 \right)^{g_2} \cdot \left( f_{\mathrm{xtalk}} \right)^{N_{\mathrm{xtalk}}} \cdot \left( f_{\mathrm{tr}} \right)^{N_{\mathrm{tr}}} \\
     \times \prod_{q \in \mathcal{Q}} \exp\left(-\frac{t_q}{T_2} \right).
    \end{aligned}
\label{eq:fidelity}
\end{equation}
  Here, $f_1$ and $f_2$ denote the fidelities of single- and two-qubit gates, respectively, while $g_1$ and $g_2$ represent their counts in the circuit. Executing two-qubit gates requires a global light field to excite atoms to the Rydberg state. However, this field can inadvertently affect non-targeted atoms within the entanglement zone, introducing crosstalk noise. The term $f_{\mathrm{xtalk}}$ accounts for the fidelity loss due to this effect, with $N_{\mathrm{xtalk}}$ representing the number of such occurrences. Furthermore, $f_{\mathrm{tr}}$ accounts for the fidelity of atom transfer operations between the SLM and AOD traps and $N_{\mathrm{tr}}$ represents the number of transfers occurring during circuit execution. Finally, the impact of decoherence is captured by the product term $\prod_{q \in \mathcal{Q}} \exp\left(- t_q / T_2\right)$ over the set of all qubits $\mathcal{Q}$. For each qubit $q$, the fidelity is reduced based on its idle time $t_q$ and the coherence time $T_2$. The idle time is the total circuit duration minus the time the qubit is actively occupied (gate execution, transport, etc.). All parameters for this model, listed in Tab.~\ref{tab:parameters}, are based on experimental demonstrations, with the qubit coherence time $T_2$ set to 1.5~seconds~\cite{bluvstein2022quantum}.

  \begin{table}[t]
  \caption{Key experimental parameters used in this work. The listed gate durations and operation fidelities are based on the values reported in~\cite{bluvstein2024logical}, while atom movement is characterized by the acceleration constraint used in our simulator.}\label{tab:parameters}
  \begin{tabular}{|p{60pt}|p{30pt}|p{70pt}|p{30pt}|}
      \hline
      Operation & Notation & Duration / Constraint & Fidelity \\
      \hline
      Single-qubit gate & $f_1$ & 52~$\mu$s & 0.9997 \\
      Two-qubit gate & $f_2$ & 0.36~$\mu$s & 0.995 \\
      Crosstalk & $f_{\mathrm{xtalk}}$ & 0.36~$\mu$s & 0.9975 \\
      Atom transfer & $f_{\mathrm{tr}}$ & 15~$\mu$s & 0.999 \\
      Atom movement & - & $a < 2.75~\mu\mathrm{m}\,\mu\mathrm{s}^{-2}$ & $\sim$1 \\
      \hline
  \end{tabular}
  \end{table}
  
  For atom movement, we assume near-unit intrinsic fidelity under sufficiently smooth transport, so its primary effect enters through additional runtime rather than through a separate direct error channel. Accordingly, the ``Atom movement'' row in Tab.~\ref{tab:parameters} reports the acceleration constraint used in our simulator rather than a fixed duration. For a displacement of length $d$, assuming a symmetric acceleration--deceleration profile with maximum acceleration $a$, we use the movement-time model
  \begin{equation}
      t_{\mathrm{mv}}(d)=2\sqrt{\frac{d}{a}}.
  \label{eq:movement_time_model}
  \end{equation}
  This model is used consistently for all compared compilers and corresponds to an acceleration-limited transport profile under the bound in Tab.~\ref{tab:parameters}. When a logical movement must be decomposed into multiple routing sub-stages, or when a short parking shift is inserted to resolve AOD conflicts, the total movement overhead is computed as the sum of the durations of all constituent movement steps.

  For a given quantum circuit, the gate counts $g_1$ and $g_2$ are predetermined by the algorithm's structure. The corresponding gate fidelities, $f_1$ and $f_2$, are primarily limited by the quality of the local laser control and intrinsic atomic properties. Consequently, the gate fidelity term in~\eqref{eq:fidelity}, $(f_1)^{g_1}\cdot(f_2)^{g_2}$, offers limited room for improvement through compilation alone.
  
  Our compilation strategy, therefore, targets the remaining dynamic factors that are highly dependent on the circuit's spatio-temporal layout. The primary optimization goals are to:
  \begin{enumerate}
      \item Minimize crosstalk events $N_\mathrm{xtalk}$ by optimizing qubit placement and gate scheduling.
      \item Reduce the number of atom transfers $N_{\mathrm{tr}}$ to limit transport-related time overhead and errors.
      \item Shorten the total circuit execution time, which directly reduces each qubit's idle time $t_q$ and mitigates decoherence.
  \end{enumerate}
  
  By targeting these factors, a compiler can significantly enhance the overall fidelity of quantum computations on neutral atom platforms.
  
\section{ZAP Compiler}\label{sec:zap}
  
Building on the distinctive features of neutral-atom quantum platforms and the fidelity analysis above, we propose \underline{ZAP}---a \underline{Z}oned \underline{A}rchitecture and \underline{P}erformant Compiler for field-programmable atom arrays. ZAP leverages the architectural separation of storage and entanglement zones to enable a deterministic, single-pass compilation process. This approach removes the primary performance bottleneck of prior methods, leading to significant gains in both compilation speed and execution fidelity. The ZAP compilation flow consists of three key steps: scheduling, mapping and placement, and routing, which are described in the following subsections.
  
\subsection{Qubit Scheduling}
  
Before execution, a quantum circuit must be transpiled into a sequence of single-qubit and CZ gates. The circuit's performance is often characterized by its depth, defined by the number of sequential \textit{stages}. Here, a stage denotes a logical scheduling block rather than a guarantee that all contained operations are physically executed simultaneously; operations with incompatible hardware requirements may be temporally ordered within the same stage.
  
ASAP scheduling is a standard technique for minimizing circuit depth. In zoned neutral-atom architectures, however, different ASAP variants interact quite differently with qubit-movement overhead. We therefore distinguish two execution policies within the same ASAP framework: a joint scheduling strategy (denoted as ASAP-joint), which assigns single- and two-qubit gates to the same logical scheduling blocks, and a separate scheduling strategy (denoted as ASAP-separate), which schedules two-qubit gates first and then fills the remaining temporal slots with single-qubit gates.

\begin{figure}[t]
    \centering
    \includegraphics[width=0.99\columnwidth]{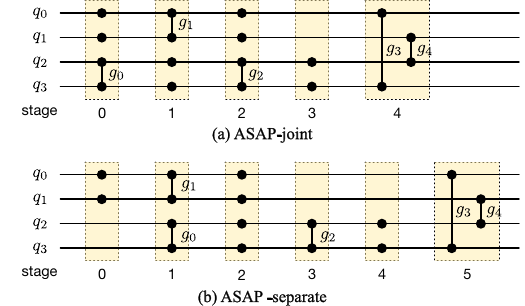}
    \caption{ASAP-joint: assigns single- and two-qubit gates to the same logical scheduling blocks to minimize logical depth. ASAP-separate: schedules two-qubit gates first, then fills single-qubit gates into available slots. This strategy may increase stage count but better matches zoned architectures by reducing shuttling overhead, and is adopted in ZAP.\label{fig:asap}}
    \label{fig:asap}
\end{figure}

\begin{figure}[t]
    \centering
    \includegraphics[width=1\linewidth]{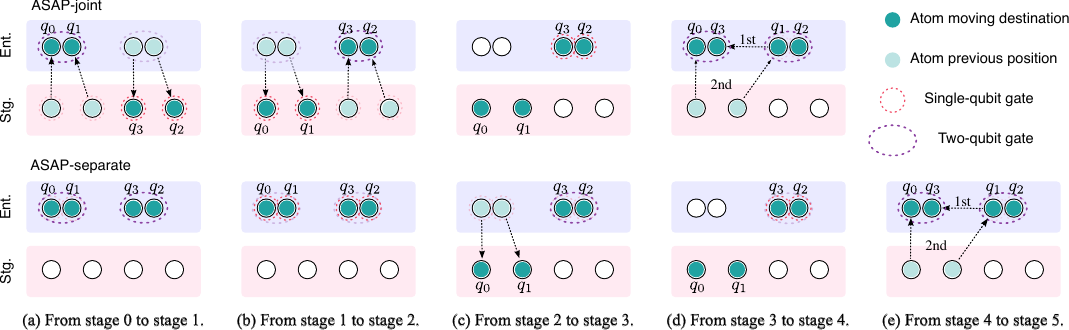}
    \caption{Step-by-step illustration of qubit movement across logical stages under the two scheduling strategies in Fig.~\ref{fig:asap}. The top row shows ASAP-joint scheduling, in which tightly coupled dependencies introduce frequent atom movements between consecutive stages. The bottom row shows ASAP-separate scheduling, in which two-qubit gates are grouped and single-qubit gates are deferred, significantly reducing inter-stage movement. Dashed arrows indicate atom transport, and highlighted regions denote gate-execution zones. Mixed stages in ASAP-joint are logical groupings; under hardware constraints, their single- and two-qubit operations remain temporally ordered.\label{fig:scheduling_exp}}
    \label{fig:scheduling_exp}
\end{figure}

The ASAP-joint strategy assigns all gate types to the earliest possible logical stages to minimize the total number of stages, as shown in Fig.~\ref{fig:asap}(a). While this approach reduces logical depth, it creates tightly coupled dependencies between consecutive stages and therefore often requires frequent qubit movement. For example, as illustrated in Fig.~\ref{fig:scheduling_exp}, ASAP-joint introduces frequent movement between consecutive stages (e.g., stages 0--1, 1--2, and 3--4) due to tightly coupled gate dependencies. In contrast, ASAP-separate significantly reduces movement by grouping two-qubit interactions and deferring single-qubit operations.
  
In contrast, ZAP adopts the ASAP-separate strategy, which first schedules all two-qubit gates and then fills the remaining temporal slots with single-qubit gates (Fig.~\ref{fig:asap}(b)). By decoupling the scheduling of gate types, this approach reduces unnecessary qubit movement and simplifies downstream placement and routing. In the same example, qubit movement is required only between stages 0 and 1, and between stages 4 and 5.

Importantly, grouping single- and two-qubit gates within the same stage does not imply simultaneous execution. In our execution model, operations within a stage are temporally ordered to respect hardware constraints. In particular, global Rydberg entangling operations and single-qubit operations that cannot be applied concurrently are serialized within the stage, whereas physically compatible operations acting on different zones can still be executed in parallel.

This design reflects an important trade-off: although ASAP-separate may increase the number of stages, it significantly reduces qubit shuttling, which is a dominant source of decoherence and operational overhead in zoned architectures. As a result, it improves overall execution efficiency despite the potentially larger circuit depth.
  
\subsection{Qubit Mapping and Placement}
  
ZAP adopts a deterministic, routing-aware placement strategy. Rather than solving a costly global optimization problem, it combines a one-time initial mapping with a lightweight stage-wise refinement that explicitly favors movement patterns likely to be executable in parallel under AOD constraints. This design preserves compilation efficiency while tightly coupling placement to the downstream routing procedure.

\textbf{Initial mapping.}
At the start of compilation, all qubits are assigned to storage sites. We first rank the storage sites according to their proximity to the entanglement zone. For each storage site $u \in \mathcal{S}$, let its average distance to the entanglement sites $\mathcal{E}$ be
\begin{equation}
    D(u) = \frac{1}{|\mathcal{E}|}\sum_{e\in \mathcal{E}}\|u-e\|,
\label{eq:storage_avg_distance}
\end{equation}
and let its nearest entanglement site be
\begin{equation}
e^\star(u)=\arg\min_{e\in \mathcal{E}}\|u-e\|,
\quad
d_{\min}(u)=\|u-e^\star(u)\|.
\label{eq:nearest_ent_site}
\end{equation}

Qubits are prioritized according to their expected importance in the circuit, with earlier stages receiving larger weights:
\begin{equation}
w(\ell)=\frac{1}{\ell+1},\quad \ell\in\{0, 1, 2, \dots, n_{\text{stage}}-1\}
\label{eq:stage_weight}
\end{equation}
where $\ell$ is the stage index, with $\ell=0$ denoting the first stage, and gate contributions decay as $\ell$ increases. The total weight for qubit $q$ is then computed as
\begin{equation}
W(q)=\sum_\ell \sum_{g\in \mathcal{G}_\ell,\; q\in g} w(\ell),
\label{eq:qubit_weight}
\end{equation}
where $\mathcal{G}_\ell$ denotes the set of gates in stage $\ell$. This weighting scheme emphasizes near-term interactions and gives higher priority to qubits that participate more frequently in earlier stages.

To further improve routing concurrency, the initial mapping incorporates a parallel-awareness term. For each qubit $q$, we identify the first stage $\ell^{\mathrm{first}}_{2q}(q)$ in which it participates in a two-qubit gate. If this first two-qubit use occurs within a short look-ahead window, then assigning $q$ to a storage site $u$ induces an estimated first movement vector from $u$ to $e^\star(u)$. We denote this vector by
\begin{equation}
v(u)=\bigl(u_x,\, e_x^\star(u),\, u_y,\, e_y^\star(u)\bigr).
\label{eq:init_vector}
\end{equation}
For a candidate site $u$, we count how many already planned vectors are incompatible with $v(u)$ under the same 2D compatibility rule later enforced by the router. Denoting this count by $c(u)$, we score each candidate site as
\begin{equation}
\mathrm{score}(u)=\lambda_{\mathrm{par}}\, c(u)+d_{\min}(u),
\label{eq:init_score}
\end{equation}
where $\lambda_{\mathrm{par}}$ controls the priority given to future parallel transport. Thus, the initial mapping balances two goals: placing important qubits near the entanglement zone and making their early movements more likely to be routed concurrently.

\begin{algorithm}[t]
    \footnotesize
    \caption{Parallel-Aware Initial Mapping}
    \label{alg:init_mapping}
    \renewcommand{\algorithmicrequire}{\textbf{Input:}}
    \renewcommand{\algorithmicensure}{\textbf{Output:}}
    \begin{algorithmic}[1]
    \Require Storage sites $\mathcal{S}$, entanglement sites $\mathcal{E}$, qubits $\mathcal{Q}$, full stage schedule $\Pi$, look-ahead window $L$, parallel weight $\lambda_{\mathrm{par}}$
    \Ensure Initial mapping $M_0$
    \ForAll{$u \in \mathcal{S}$}
        \State Compute $D(u)$ using Eq.~\eqref{eq:storage_avg_distance}
        \State Compute $e^\star(u)$ and $d_{\min}(u)$ using Eq.~\eqref{eq:nearest_ent_site}
    \EndFor
    \State $\mathcal{S}_{\mathrm{sorted}} \gets$ storage sites sorted by increasing $D(u)$
    \ForAll{$q \in \mathcal{Q}$}
        \State Compute $W(q)$ using Eq.~\eqref{eq:qubit_weight}
        \State Determine $\ell^{\mathrm{first}}_{2q}(q)$
    \EndFor
    \State $\mathcal{Q}_{\mathrm{sorted}} \gets$ qubits sorted by decreasing $W(q)$
    \State $M_0 \gets \emptyset$, $\mathcal{V}_{\mathrm{plan}} \gets \emptyset$
    \ForAll{$q \in \mathcal{Q}_{\mathrm{sorted}}$}
        \ForAll{unassigned site $u \in \mathcal{S}_{\mathrm{sorted}}$}
            \If{$\ell^{\mathrm{first}}_{2q}(q) \neq \bot$ \textbf{and} $\ell^{\mathrm{first}}_{2q}(q) \le L$}
                \State Form vector $v(u)$ using Eq.~\eqref{eq:init_vector}
                \State $c(u) \gets$ number of vectors in $\mathcal{V}_{\mathrm{plan}}$ that are AOD-incompatible with $v(u)$ under the router's row/column compatibility rule
            \Else
                \State $c(u) \gets 0$
            \EndIf
            \State $\mathrm{score}(u) \gets \lambda_{\mathrm{par}}\,c(u)+d_{\min}(u)$
        \EndFor
        \State $u^\star \gets \arg\min_u \mathrm{score}(u)$
        \State Assign $M_0(q) \gets u^\star$
        \If{$\ell^{\mathrm{first}}_{2q}(q) \neq \bot$ \textbf{and} $\ell^{\mathrm{first}}_{2q}(q) \le L$}
            \State Add $v(u^\star)$ to $V_{\mathrm{plan}}$
        \EndIf
    \EndFor
    \State \Return $M_0$
    \end{algorithmic}
\end{algorithm}
  
\textbf{Stage-wise placement.}
After the initial mapping, the placer updates the qubit layout stage by stage. For each stage, it first handles qubits that remain in the entanglement zone but are not active in the current two-qubit stage. Unlike prior zoned compilers that typically move idle qubits back to the storage zone after two-qubit operations, ZAP dynamically decides on a per-qubit and per-stage basis whether an idle qubit should remain in the entanglement zone or be returned to storage. This decision is governed by an explicit trade-off between leaving an idle qubit in the entanglement zone, which may expose it to repeated global two-qubit illumination, and moving it back to storage, which incurs atom-transfer and decoherence costs.

For an idle qubit $q$ at stage $\ell$, let $\ell_{\mathrm{next}}(q)$ denote the next stage in which $q$ participates in a two-qubit gate. If no such stage exists, we set $\ell_{\mathrm{next}}(q)$ to the end of the circuit. Let $k_{q,\ell}$ denote the number of stages during which $q$ would remain exposed to two-qubit-zone crosstalk if it stays in place. The resulting crosstalk loss is estimated as
\begin{equation}
L^{\mathrm{xtalk}}_{q,\ell}
=
k_{q,\ell}\bigl(-\ln f_{\mathrm{2q,idle}}\bigr),
\label{eq:idle_crosstalk_cost}
\end{equation}
where $f_{\mathrm{2q,idle}}$ is the effective fidelity assigned to an idle qubit during a two-qubit stage. If $q$ is moved back to storage, the compiler instead incurs a transfer loss
\begin{equation}
L^{\mathrm{tr}}_{q,\ell}
=
n_{\mathrm{tr}}(q,\ell)\bigl(-\ln f_{\mathrm{tr}}\bigr),
\label{eq:idle_transfer_cost}
\end{equation}
where $n_{\mathrm{tr}}(q,\ell)=2$ if $q$ is never used again in a two-qubit gate and $n_{\mathrm{tr}}(q,\ell)=4$ otherwise. In addition, moving the qubit introduces extra runtime and hence additional decoherence. Let $d_{q,\ell}$ be the distance from its current location to the selected storage site, and let $t_{\mathrm{mv}}(d_{q,\ell})$ defined by Eq.~\eqref{eq:movement_time_model} denote the movement duration model. The decoherence term is approximated as
\begin{equation}
L^{\mathrm{dec}}_{q,\ell}
\approx
\alpha_{\mathrm{idle}} \frac{(N_q-1)\,\tilde{t}_{q,\ell}}{T_2},
\label{eq:idle_decoherence_cost}
\end{equation}
where $N_q$ is the total number of qubits, $T_2$ is the coherence time, $\alpha_{\mathrm{idle}}$ is a calibration weight, and $\tilde{t}_{q,\ell}$ is the estimated additional movement time associated with sending the qubit out and, if needed, bringing it back later. The placement decision is then
\begin{equation}
x^\star_{q,\ell}
=
\begin{cases}
1, & \text{if } L^{\mathrm{xtalk}}_{q,\ell} > L^{\mathrm{tr}}_{q,\ell}+L^{\mathrm{dec}}_{q,\ell},\\
0, & \text{otherwise},
\end{cases}
\label{eq:idle_decision}
\end{equation}
where $x^\star_{q,\ell}=1$ means that $q$ is returned to storage.

After idle-qubit management, the placer assigns target entanglement sites for the active two-qubit gates of the current stage. If one qubit of a gate already occupies one site of a valid entanglement pair, the other qubit is moved to the partner site whenever possible. Otherwise, the placer selects a free entanglement pair. When multiple free entanglement pairs are available, ZAP deterministically selects the pair that minimizes the routing-aware placement score in Eq.~\eqref{eq:pair_move_score}. This lightweight greedy heuristic favors physically shorter movements while simultaneously reducing anticipated AOD conflicts in the subsequent routing stage.

Concretely, if qubit $q$ is moved to site $p$, the placer forms its movement vector and counts incompatibilities with previously committed vectors in the same stage. For a single move, the score takes the form
\begin{equation}
\mathrm{score}_{\mathrm{single}}(q,p)
=
\lambda_{\mathrm{par}}\,
\mathrm{conflicts}(q,p)
+
\|M(q)-p\|,
\label{eq:single_move_score}
\end{equation}
and for a two-qubit placement onto an ordered pair $(p_0,p_1)$ the score becomes
\begin{equation}
\begin{aligned}
\mathrm{score}_{\mathrm{pair}}(q_0,q_1,p_0,p_1)
=
\lambda_{\mathrm{par}}\,
\mathrm{conflicts}(q_0,q_1,p_0,p_1) \\
+
\|M(q_0)-p_0\|
+
\|M(q_1)-p_1\|.
\end{aligned}
\label{eq:pair_move_score}
\end{equation}
Thus, stage-wise placement is not a purely geometric nearest-pair assignment. Instead, it incrementally commits qubit movements in a manner that preserves future routing concurrency under AOD constraints.

\begin{algorithm}[t]
    \footnotesize
    \caption{Routing-Aware Stage-Wise Placement}
    \label{alg:stage_placement}
    \renewcommand{\algorithmicrequire}{\textbf{Input:}}
    \renewcommand{\algorithmicensure}{\textbf{Output:}}
    \begin{algorithmic}[1]
    \Require Current mapping $M$, current stage $\ell$, stage gates $\mathcal{G}_\ell$, full schedule $\Pi$, storage sites $\mathcal{S}$, entanglement pairs $\mathcal{P}$
    \Ensure Target mapping $M_{\mathrm{target}}$ for stage $\ell$
    \State $M_{\mathrm{target}} \gets M$
    \State $V_{\mathrm{plan}} \gets \emptyset$
    \State \textbf{Step 1: idle-qubit management}
    \ForAll{qubit $q$ currently in the entanglement zone}
        \If{$q$ is not active in a two-qubit gate at stage $\ell$}
            \State Find $\ell_{\mathrm{next}}(q)$
            \State Compute $L^{\mathrm{xtalk}}_{q,\ell}$ using Eq.~\eqref{eq:idle_crosstalk_cost}
            \State Select the best candidate storage site
            \State Compute $L^{\mathrm{tr}}_{q,\ell}$ and $L^{\mathrm{dec}}_{q,\ell}$ using Eq.~\eqref{eq:idle_transfer_cost} and Eq.~\eqref{eq:idle_decoherence_cost}
            \If{$L^{\mathrm{xtalk}}_{q,\ell} > L^{\mathrm{tr}}_{q,\ell}+L^{\mathrm{dec}}_{q,\ell}$}
                \State Move $q$ to storage and commit its movement vector
            \EndIf
        \EndIf
    \EndFor
    \State \textbf{Step 2: active two-qubit placement}
    \ForAll{two-qubit gate $(q_0,q_1)$ in stage $\ell$}
        \If{$q_0$ and $q_1$ are already on a valid entanglement pair}
            \State Continue to the next gate
        \ElsIf{$q_0$ already occupies one site of a partially filled valid pair}
            \State Move $q_1$ to the partner site using the lowest AOD-aware routing score
            \State Commit all newly selected movement vectors
        \ElsIf{$q_1$ already occupies one site of a partially filled valid pair}
            \State Move $q_0$ to the partner site using the lowest AOD-aware routing score
            \State Commit all newly selected movement vectors
        \Else
            \State Choose a free entanglement pair $(p_0,p_1)\in\mathcal{P}$
            \State Orient $(p_0,p_1)$ to minimize the AOD-aware score in Eq.~\eqref{eq:pair_move_score}
            \State Assign $q_0\to p_0$ and $q_1\to p_1$
            \State Commit all newly selected movement vectors
        \EndIf
    \EndFor
    \State \Return $M_{\mathrm{target}}$
    \end{algorithmic}
\end{algorithm}

\subsection{Qubit Routing}\label{subsec:qubit_routing}

\begin{figure}[t]
    \centering
    \includegraphics[width=0.99\columnwidth]{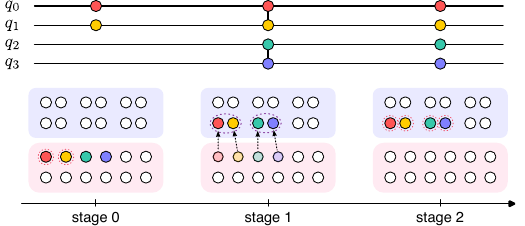}
    \caption{Routing of a quantum circuit in ZAP. The circuit is divided into sequential stages, each containing gates that can be executed in parallel. In stage 0, single-qubit gates are applied to $q_0$ and $q_1$ in the storage zone. To execute the two-qubit gates in stage 1, qubits $q_0$ through $q_3$ are shuttled to the entanglement zone. In stage 2, subsequent single-qubit gates are applied. This dynamic shuttling between zones, dictated by the circuit structure, realizes the complete routing of the quantum circuit.\label{fig:routing_overview}}
    \label{fig:routing_overview}
\end{figure}

Once the target placement for a stage has been determined, the router generates the physical instruction sequence that transforms the current qubit configuration into that target layout. Routing therefore bridges consecutive logical stages: it realizes the qubit movements implied by stage-wise placement and ensures that the required qubits reach valid physical locations before gate execution begins.

Figure~\ref{fig:routing_overview} illustrates the stage-level role of routing in ZAP. The circuit is executed as a sequence of stages: single-qubit stages can often be completed directly in the storage zone, whereas two-qubit stages require selected qubits to be transported into the entanglement zone and aligned with valid interaction pairs. In the example, stage 0 applies single-qubit gates to $q_0$ and $q_1$ without movement; stage 1 shuttles $q_0$ through $q_3$ into the entanglement zone to execute two two-qubit gates; and the final stage again applies single-qubit gates locally. Routing is therefore not an isolated geometric subroutine, but the mechanism that transforms the current qubit layout into the target layout required by the next stage.

For a given stage, let $M$ denote the current mapping and let $M_{\mathrm{target}}$ denote the target mapping produced by the placer. Each qubit whose source and destination differ induces a movement vector
\begin{equation}
    v_q = \bigl(x_q^{\mathrm{src}},\, x_q^{\mathrm{dst}},\, y_q^{\mathrm{src}},\, y_q^{\mathrm{dst}}\bigr).
\label{eq:routing_vector}
\end{equation}
The full routing task is therefore represented by the set
\begin{equation}
\mathcal{V}=\{v_q \mid M(q)\neq M_{\mathrm{target}}(q)\}.
\label{eq:routing_set}
\end{equation}

This transport is not executed as one monolithic movement. Instead, the router repeatedly extracts a subset of safe, mutually compatible moves, executes that batch, updates the current mapping, and then continues with the remaining vectors. To avoid immediate target-site contention, the router first identifies the set of \emph{safe} vectors,
\begin{equation}
\mathcal{V}_{\mathrm{safe}} = \left\{v\in \mathcal{V} \;\middle|\; \mathrm{target}(v)\notin \mathcal{P}_{\mathrm{start}}
\right\},
\label{eq:safe_vectors}
\end{equation}
where $\mathcal{P}_{\mathrm{start}}$ denotes the set of source positions of all unfinished moves. If $\mathcal{V}_{\mathrm{safe}}$ is empty, the router executes a remaining move to break the cycle.

Among the safe vectors, the router determines which moves are compatible under the AOD row/column constraint. Two vectors are compatible only if they preserve row and column ordering and do not request conflicting actuation on the same source or destination line. The router then builds a compatibility graph on $\mathcal{V}_{\mathrm{safe}}$ and selects a greedy maximal independent set as the next routing batch.

Not all conflicts require full serialization. Figure~\ref{fig:parking_strategy} illustrates a representative routing example in which two atoms start in the same row but must move to different target rows. Rather than serializing the two moves, ZAP first activates both atoms, applies a short parking displacement to the upper atom, and then performs the main movement in parallel. After the atoms reach their destinations, they are deactivated. This example shows how a small intermediate adjustment can convert an otherwise incompatible movement pattern into a parallel one.

More generally, parking operations are inserted whenever direct transport is blocked by AOD selection ambiguity or by geometric interference with static atoms. These auxiliary displacements are explicit parts of the routing instruction stream, and their runtime overhead is included in the physical execution schedule. After each batch is completed, the corresponding vectors are removed from $\mathcal{V}$, the mapping is updated, and the process repeats until all movements are finished. ZAP routing is therefore an iterative safe-batch transport procedure with exact compatibility checking and parking-based local conflict resolution.

\begin{figure}[t]
    \centering
    \includegraphics[width=0.99\columnwidth]{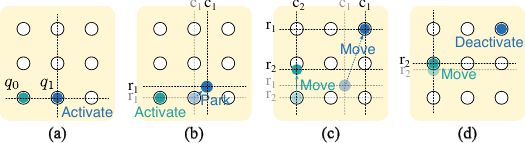}
    \caption{Instruction-level example of parking-enabled parallel transport. (a) Two atoms start in the same row and are activated. (b) A short parking displacement is applied to the upper atom to resolve the initial AOD row/column conflict. (c) The two atoms then execute their main movement in parallel toward different target rows. (d) After reaching their destinations, the atoms are deactivated. This example illustrates how ZAP uses a brief parking step to convert an initially incompatible configuration into one that supports concurrent transport.\label{fig:parking_strategy}}
    \label{fig:parking_strategy}
\end{figure}

\section{Evaluation Setup}\label{sec:setup}

To comprehensively evaluate the performance of ZAP, we develop a simulation framework implemented in Python 3.12 and executed on a system equipped with an Apple M4 processor and 16~GB of memory. Our evaluation adopts experimentally realistic parameters based on state-of-the-art neutral-atom platforms and compares ZAP against representative baseline compilers.

\subsection{Hardware Parameter Setting}

\begin{figure}[t]
    \centering
    \includegraphics[width=0.99\columnwidth]{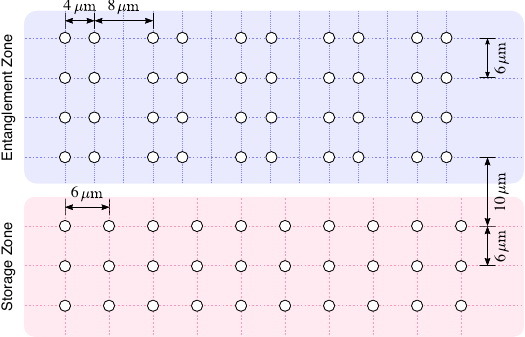}
    \caption{The storage zone consists of atoms arranged with a spacing of 6 $\mu$m in both the $x$- and $y$-directions, where qubits are held when not participating in two-qubit gate operations. The entanglement zone features atom pairs with an intra-pair spacing of 4 $\mu$m and a minimum inter-pair spacing of 6 $\mu$m, facilitating controlled interactions for quantum entanglement operations. For commonly used qubit atoms, the Rydberg blockade radius is typically 5 $\mu$m. For example, the blockade radius is approximately 4.3 $\mu$m for Rb, which informs the atom arrangement in our simulation.\label{fig:platform}}
    \label{fig:platform}
\end{figure}
  
For rubidium (Rb) atoms, the Rydberg blockade radius is approximately 4.3~$\mu$m~\cite{bluvstein2024logical}. Accordingly, as shown in Fig.~\ref{fig:platform}, the atomic spacing in the storage zone was set to 6~$\mu$m in both the $x$ and $y$ directions. This spacing ensures that atoms in the Rydberg state do not interact with their neighbors and allows each atom to perform single-qubit gates without interference. In the entanglement zone, the spacing between atoms within each qubit pair used for two-qubit gates along the $x$-direction was set to 4~$\mu$m, while a minimum separation of 6~$\mu$m was maintained between qubit pairs at different sites. This configuration enables parallel entangling operations because each active pair lies within the Rydberg blockade radius while different pairs remain sufficiently separated. To prevent interference between the laser systems in the storage and entanglement zones, a separation of 10~$\mu$m was maintained between the two regions. 

\subsection{Benchmarks}

To comprehensively evaluate the performance and scalability of the ZAP compiler, we consider two categories of quantum circuits: structured algorithmic benchmarks and synthetic random circuits. The benchmark suite spans a broad range of system sizes, including small (2--10 qubits), medium (11--27 qubits), and large (28 or more qubits) instances.

\subsubsection{Structured Quantum Algorithms}

We selected a diverse set of representative quantum algorithms to assess the compiler across different logical structures. The suite includes four classes of workloads: (1) arithmetic and mathematical transforms, including Adder, Multiplier, and the Quantum Fourier Transform (QFT), which often exhibit dense, hierarchical entanglement patterns such as the butterfly network in QFT; (2) optimization and simulation circuits, including the Quantum Approximate Optimization Algorithm (QAOA), Ising models, and Max-Cut (SAT) solvers, which typically exhibit regular grid-like or near-neighbor connectivity; (3) data-structure and machine-learning circuits, including Quantum Random Access Memory (QRAM), K-Nearest Neighbors (KNN), and Variational Quantum Classifiers (VQC), with QRAM notably featuring a complex tree-like topology; and (4) state-preparation circuits, including GHZ, W-state, and Cat-state circuits, which provide high-entanglement-depth benchmarks for fidelity scaling.

These workloads allow us to examine how different compiler heuristics navigate architecture-specific constraints, particularly the trade-off between atom transport and crosstalk in irregular interaction patterns such as QRAM.

\subsubsection{Random Circuit Generation}

To evaluate scalability without bias toward any specific algorithmic motif, we generate synthetic random circuits from 3-regular interaction graphs. For each tested qubit count $N$, we use the NetworkX function \texttt{nx.random\_regular\_graph(d=3, n=N)} to construct an undirected graph $G=(V,E)$, where each vertex $v\in V$ represents a logical qubit and each edge $(u,v)\in E$ corresponds to one logical two-qubit CZ interaction between qubits $u$ and $v$. Consequently, every qubit participates in exactly three entangling interactions, and each random instance contains $|E|=3N/2$ two-qubit gates.

We choose 3-regular graphs for two reasons. First, fixing the degree at $d=3$ keeps the interaction density sparse and consistent across problem sizes, allowing the compiler to be evaluated under a controlled scaling regime rather than under varying gate densities. Second, random regular graphs suppress benchmark-specific structural artifacts, such as hubs, chains, and tree-like motifs, that could favor one heuristic over another. As a result, these circuits provide a clean average-case benchmark for studying how compilation time, execution time, movement overhead, and fidelity scale with the number of qubits. In this sense, the random-circuit study isolates scalability with respect to system size, whereas the structured benchmarks capture behavior on application-specific connectivity patterns.

\subsection{Baseline}

We compare ZAP against three representative neutral-atom compilers: Enola, ZAC, and PowerMove. Enola targets non-zoned architectures and leverages graph-coloring-based scheduling to minimize circuit depth, thereby serving as a baseline that highlights the advantage of zoned architectures in suppressing crosstalk~\cite{tan2024compilation}. ZAC is a zoned neutral-atom compiler that employs iterative optimization (e.g., simulated annealing) for initial mapping and qubit reuse, and thus provides a baseline for evaluating the trade-off between compilation quality and runtime~\cite{lin2025reuse}. PowerMove is a strong recent compiler that integrates multi-stage heuristics for joint scheduling and routing, making it a strong heuristic baseline for zoned architectures~\cite{ruan2024powermove}.

Since single-qubit gate implementations are identical across the compared compilers and their counts are not affected by compilation strategy, we exclude single-qubit gate fidelity from the evaluation metric to focus on the dominant sources of variation. Specifically, we evaluate fidelity in terms of two-qubit gates, atom transfer, decoherence, and crosstalk. Nevertheless, single-qubit gates are fully accounted for during compilation and scheduling and are included in the optimization process.

\section{Evaluation Results}\label{sec:results}

We evaluate ZAP on structured benchmarks and random circuits in terms of fidelity, circuit execution time, and compilation time relative to prior neutral-atom compilers, and then use ablation studies to quantify the impact of its main design choices.

\subsection{Fidelity Performance}

\begin{figure*}[t]
    \centering
    \includegraphics[width=1\linewidth]{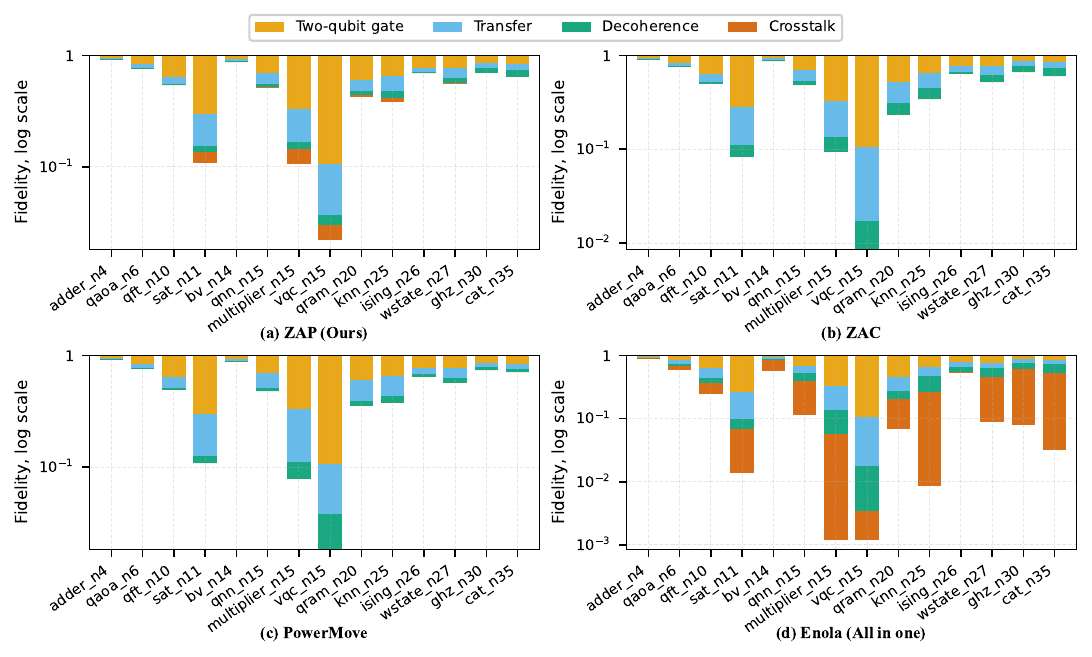}
    \caption{Structured-benchmark fidelity breakdown for four compilers: (a) ZAP, (b) ZAC, (c) PowerMove, and (d) Enola. Each stacked bar shows the remaining fidelity after accounting for two-qubit-gate error, atom-transfer error, decoherence, and crosstalk for one benchmark circuit. This figure highlights how different compilers trade off transport, decoherence, and crosstalk on the same workload suite.\label{fig:fidelity_detailed}}
    \label{fig:fidelity_detailed}
\end{figure*}

\begin{figure*}[t]
    \centering
    \includegraphics[width=1\linewidth]{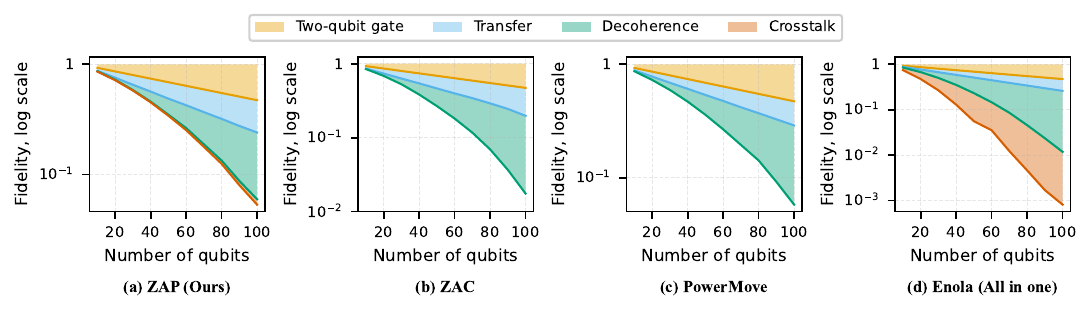}
    \caption{Random-circuit fidelity scaling on 3-regular graph benchmarks for (a) ZAP, (b) ZAC, (c) PowerMove, and (d) Enola. Each benchmark is constructed by generating a random 3-regular interaction graph and instantiating one CZ gate per edge. For each compiler, the colored regions show the cumulative fidelity degradation induced by two-qubit gates, atom transfer, decoherence, and crosstalk as the number of qubits increases from 10 to 100. \label{fig:fidelity_detailed_random}}
    \label{fig:fidelity_detailed_random}
\end{figure*}

To assess ZAP's fidelity performance, we estimate circuit fidelity under the hardware parameters described above on a benchmark suite spanning a range of circuit depths and problem sizes. Here, depth denotes the number of stages in the ASAP-separate schedule. The suite includes small (2 to 10 qubits), medium (11 to 27 qubits), and large (28 or more qubits) instances, enabling a broad evaluation of both performance and scalability.

We first examine the fidelity budget of the compiled circuits, with the structured-benchmark breakdown shown in Fig.~\ref{fig:fidelity_detailed} and the random-circuit scaling study shown in Fig.~\ref{fig:fidelity_detailed_random}. These analyses include the main loss channels relevant to compilation: two-qubit-gate error, atom transfer between the SLM and AOD, decoherence, and crosstalk. Among these terms, the gate contribution is largely hardware-determined and therefore serves as a common background factor, whereas atom transfer, decoherence, and crosstalk depend strongly on compilation decisions. Across both structured and random benchmarks, ZAP's zoned architecture and parallel transport strategy are particularly effective at suppressing crosstalk and controlling transfer overhead, while decoherence gradually becomes the dominant residual bottleneck as circuit size increases.

An important observation from Fig.~\ref{fig:fidelity_detailed_random} is that the fidelity curves of ZAP and PowerMove are much closer on random circuits than on structured algorithmic benchmarks. This behavior is expected. The random instances are generated from 3-regular graphs, so for a given qubit count each compiler sees the same sparse interaction density and the same total number of two-qubit interactions ($3N/2$). Because these graphs are topologically homogeneous and lack high-contrast motifs such as the butterfly structure in QFT or the tree-like reuse pattern in QRAM, they provide fewer opportunities for a compiler to gain a large advantage from specialized scheduling or placement decisions. As a result, hardware-determined gate and decoherence terms dominate, while compiler-dependent differences in movement, reuse, and crosstalk tend to average out.

By contrast, structured algorithms amplify the effects of compiler design because they contain nonuniform connectivity patterns and repeated motifs that interact strongly with movement policies, qubit reuse, and zone transitions. In these workloads, differences in scheduling and placement quality accumulate systematically across stages, which is why the fidelity gaps in Fig.~\ref{fig:fidelity_detailed} are more pronounced than in the random-circuit study.

\begin{figure}[t]
    \centering
    \includegraphics[width=0.99\columnwidth]{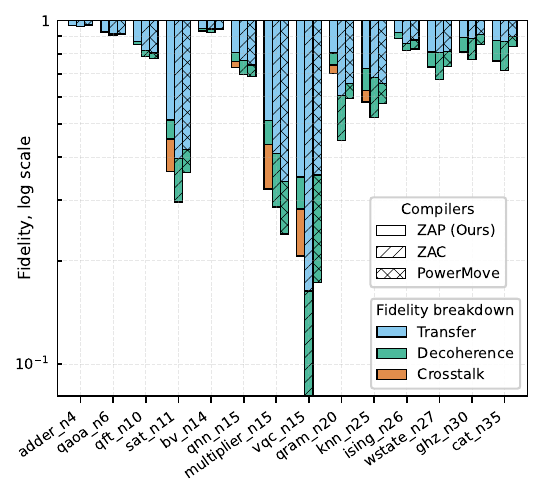}
    \caption{Direct comparison of compiler-dependent fidelity losses on the structured benchmark suite. For each benchmark, three adjacent bars correspond to ZAP, ZAC, and PowerMove, distinguished by hatch pattern. The stacked components show only the compiler-dependent loss channels---atom transfer, decoherence, and crosstalk---while gate-infidelity terms shared across compilers are omitted for clarity. This visualization isolates the source of ZAP's fidelity advantage over other zoned compilers.\label{fig:fidelity}}
    \label{fig:fidelity}
\end{figure}

Our primary finding is that ZAP consistently outperforms the Enola baseline on both structured and random benchmarks. In the structured suite of Fig.~\ref{fig:fidelity_detailed}, Enola exhibits the largest crosstalk-induced degradation, whereas ZAP preserves visibly higher fidelity across most workloads. The same qualitative trend appears in the random-circuit study of Fig.~\ref{fig:fidelity_detailed_random}, although the gap is smaller because 3-regular instances are more topologically homogeneous and therefore offer less room for compiler-dependent differentiation.

Relative to the stronger zoned baselines ZAC and PowerMove, ZAP's advantage is more workload-dependent but remains clear on the benchmarks that best expose compiler quality. As shown in Figs.~\ref{fig:fidelity_detailed} and~\ref{fig:fidelity}, ZAP is typically best or near-best on structured workloads with irregular connectivity and nonuniform reuse, where scheduling, placement, and transport decisions accumulate over many stages. A representative example is qft\_n10, where ZAP achieves a fidelity of 0.541 compared with 0.494 for PowerMove. By contrast, on more regular state-preparation circuits such as ghz\_n30 and cat\_n35, the gaps among zoned compilers are smaller because the movement pattern is simpler and leaves less room for compiler-level differentiation. This contrast reinforces the main point: ZAP's largest fidelity gains arise not from changing the hardware error model, but from making better trade-offs among crosstalk, transport, and waiting time on structurally challenging workloads.

A more fine-grained reading of the fidelity breakdown reveals that different benchmark families stress different compiler decisions. For state-preparation circuits such as GHZ and Cat, the interaction pattern is highly regular and the reuse opportunities are easy to identify, so all zoned compilers exhibit relatively similar crosstalk behavior and the remaining differences mainly reflect movement and decoherence overhead. For arithmetic and transform circuits such as Adder and QFT, long-range and repeated interaction patterns amplify the cost of unnecessary shuttling, making transfer and decoherence much more visible in the fidelity budget. For optimization and simulation benchmarks such as Ising and QAOA, the interaction structure is more localized and regular, so the gaps among zoned compilers are moderate but still reflect how effectively each method controls movement accumulation over many stages. The clearest separation appears in irregular data-access and machine-learning-style workloads such as QRAM and KNN, where nonuniform connectivity and qubit reuse patterns interact strongly with zone transitions. In those cases, ZAP benefits most from its cost-aware idle-qubit management and routing-aware placement, which reduce avoidable return-and-reload cycles and thereby lower both transport-related loss and movement-induced decoherence.

This interpretation is reinforced by the direct comparison of compiler-dependent error channels in Fig.~\ref{fig:fidelity}. There, only transfer, decoherence, and crosstalk losses are shown, allowing the influence of compiler decisions to be isolated from gate-infidelity terms common to all methods. The figure shows that ZAP almost completely suppresses crosstalk on many benchmarks while keeping transfer and decoherence losses competitive with, and often slightly below, those of ZAC and PowerMove. Together with the random-circuit trends in Fig.~\ref{fig:fidelity_detailed_random}, this indicates that ZAP's fidelity advantage comes not from changing the underlying hardware error model, but from making better scheduling, placement, and routing decisions under the same hardware constraints.
  
While ZAP significantly reduces compilation-dependent errors, decoherence remains largely hardware-limited. Improvements in vacuum quality, magnetic-field stability, laser-noise suppression, and coherence-preserving control techniques can further extend the effective coherence time of neutral-atom qubits. As these hardware-level limitations are relaxed, the relative impact of compilation choices on crosstalk and transport overhead will become even more pronounced. By effectively managing these compilation-level errors, ZAP is well-positioned to maximize the potential of future-generation neutral-atom hardware.

\subsection{Circuit Execution Time}

\begin{figure*}[t]
    \centering
    \includegraphics[width=1\linewidth]{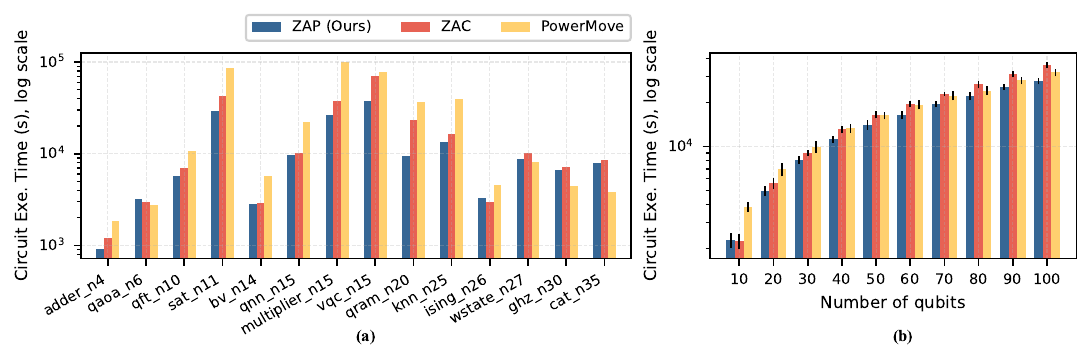}
    \caption{Circuit execution time comparison for zoned compilers. (a) Execution time on the structured benchmark suite. (b) Execution time versus qubit count for random 3-regular circuits, with one CZ interaction instantiated per graph edge. We report the physical makespan of the compiled schedule, including gate durations, SLM--AOD transfers, in-array movement, and extra delays inserted to resolve AOD conflicts. Unlike compilation time, execution time is governed mainly by shared hardware primitives, so the gap among ZAP, ZAC, and PowerMove is more modest; nevertheless, ZAP remains consistently competitive and is often the fastest due to reduced shuttling overhead.\label{fig:execution_time}}
    \label{fig:execution_time}
\end{figure*}

We next compare the physical execution time of the compiled circuits, shown in Fig.~\ref{fig:execution_time}. For each compiled schedule, we report the hardware-level makespan, including two-qubit gate durations, SLM--AOD transfer time, in-array transport time, and the additional waiting time introduced by AOD row/column conflict resolution and path-clearing constraints. This metric is directly relevant to fidelity because a longer makespan increases the circuit's exposure to decoherence. We restrict this comparison to zoned baselines, since ZAC, PowerMove, and ZAP share the same storage--entanglement execution model; Enola targets a different, non-zoned architecture, making a like-for-like physical-runtime comparison less meaningful.

The gap in Fig.~\ref{fig:execution_time} is naturally smaller than the gap in classical compilation time because ZAP, ZAC, and PowerMove all operate under the same underlying hardware primitives. In other words, no compiler can arbitrarily change the native gate speed or the atom-transport velocity. A fair runtime comparison therefore asks a more focused question: given the same hardware parameters, which compiler introduces less unnecessary movement and less scheduling overhead in the final executable schedule?

Under this criterion, ZAP performs favorably. Across both structured and random benchmarks, ZAP remains consistently competitive and is often the fastest among the zoned compilers. The advantage is particularly clear on larger and more irregular workloads, where ZAP's ASAP-separate scheduling, cost-aware idle-qubit management, and routing-aware placement reduce avoidable return-and-reload cycles and shorten the total transport path before each entangling stage. Consequently, ZAP improves fidelity by suppressing crosstalk without paying for that benefit with longer execution time.

\subsection{Classical Compilation Efficiency}

\subsubsection{Compilation Speedup}
  
Compilation time is a major system-level bottleneck in the quantum computing stack because it directly affects workflow throughput and the practicality of iterative methods. ZAP addresses this bottleneck through compiler--architecture co-design for a partitioned neutral-atom platform. In particular, ZAP uses a deterministic, single-pass compilation flow that maps logical qubits to physical atoms without the iterative search procedures used in ZAC and Enola and without the multi-stage co-optimization framework used in PowerMove.

\begin{figure*}[t]
    \centering
    \includegraphics[width=1\linewidth]{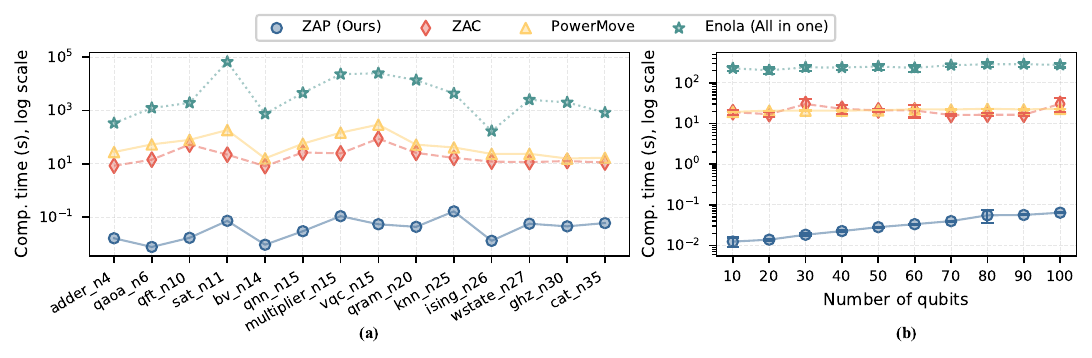}
    \caption{Compilation time comparison for ZAP, ZAC, Enola, and PowerMove. (a) Compilation time for the suite of structured quantum algorithms. (b) Compilation time as a function of qubit count for randomly generated circuits. ZAP exhibits substantially lower compilation time and better scaling.\label{fig:compilation_time}}
    \label{fig:compilation_time}
\end{figure*}
  
We first compare ZAP with ZAC, PowerMove, and Enola on the structured benchmarks used in the fidelity study. As shown in Fig.~\ref{fig:compilation_time}(a), ZAP delivers a substantial performance advantage across the entire suite. Relative to ZAC and PowerMove, compilation time is reduced from tens of seconds to below 0.1~s on many benchmarks, corresponding to consistent two- to three-order-of-magnitude speedups. The gap relative to Enola is even larger, reaching four to six orders of magnitude on this suite. These results show that ZAP largely removes compilation as a practical bottleneck and enables interactive or iterative compilation workflows.

To verify that this advantage is not specific to structured algorithms, we also evaluate randomly generated circuits of increasing size. Figure~\ref{fig:compilation_time}(b) shows that ZAP remains below 0.1~s even up to 100 qubits, whereas ZAC and PowerMove scale into the tens of seconds. This corresponds to speedups of roughly 100$\times$ to 1,000$\times$ and indicates that ZAP's efficiency stems from its underlying compilation strategy rather than from workload-specific structure.

\begin{figure*}[t]
    \centering
    \includegraphics[width=1\linewidth]{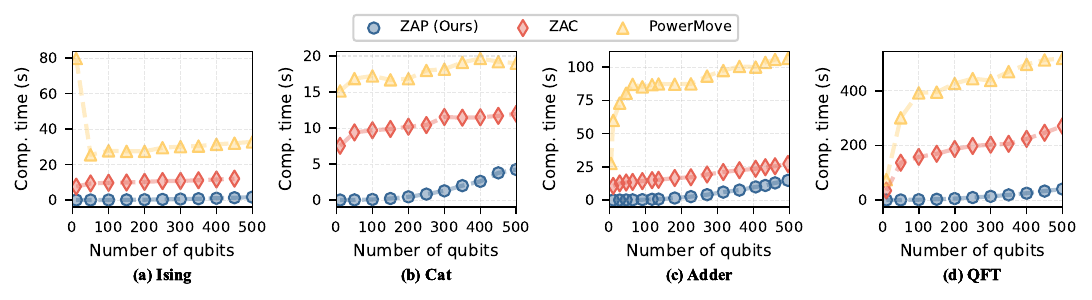}
    \caption{Compilation time scalability comparison between ZAP, ZAC, and PowerMove for four representative algorithms. From left to right, the panels correspond to (a) Ising, (b) Cat, (c) Adder, and (d) QFT. Each panel plots compilation time versus the number of qubits up to 500. Enola is omitted from this comparison due to its substantially larger compilation overhead, as established in Fig.~\ref{fig:compilation_time}. Across all four workloads, ZAP consistently exhibits both the lowest absolute compilation time and the flattest growth trend.\label{fig:results_runtime}}
    \label{fig:results_runtime}
\end{figure*}

Fast compilation is only useful if it also scales to large problem sizes. We therefore evaluate ZAP on four representative algorithms---Ising, Cat, Adder, and QFT---up to 500 qubits (Fig.~\ref{fig:results_runtime}), omitting Enola because its compilation overhead is already substantially larger in Fig.~\ref{fig:compilation_time}. The results show that ZAP's performance advantage persists at scale. Even at 500 qubits, ZAP remains orders of magnitude faster than both ZAC and PowerMove. Notably, ZAP compiles a 500-qubit QFT circuit in under a minute, whereas the iterative search procedures in ZAC and the heavier co-optimization flow in PowerMove become computationally prohibitive. This scalability is not merely an incremental improvement; it is a key enabler for exploring large-scale quantum algorithms that are difficult to study with iterative compilation methods. 

\subsubsection{Complexity Analysis}

The divergence in compilation scaling observed in Fig.~\ref{fig:compilation_time} and Fig.~\ref{fig:results_runtime} reflects the underlying algorithmic structure of each framework.

ZAC follows a staged heuristic pipeline rather than a single monolithic optimizer. Its runtime arises from several coupled subproblems, including gate scheduling, simulated-annealing-based initial placement, intermediate placement via minimum-weight bipartite matching, and routing under movement-conflict constraints using MIS-style selection. Consequently, its empirical complexity depends strongly on topology and geometry. When interaction graphs are locally bounded and structurally homogeneous, as in random 3-regular instances, conflict patterns are more regular and scaling is typically smoother. For structured circuits with long-range and highly nonuniform dependencies, such as QFT-like patterns, conflict graphs become denser and less regular, leading to larger runtime variability.

PowerMove also adopts a multi-stage heuristic flow, combining gate-stage formation, simulated-annealing-based initial placement, and movement scheduling with conflict-aware grouping. Relative to a simpler single-pass compiler, this introduces additional optimization stages and therefore higher runtime overhead. As the number of qubits and the interaction depth increase, the total optimization and scheduling workload grows accordingly.

In contrast, ZAP leverages the structural constraints of the zoned architecture to convert a potentially intractable search problem into a deterministic, single-pass process. By partitioning the array into storage and entanglement zones, ZAP reduces placement and routing to a sequence of greedy look-ahead decisions.

To make this statement more precise, let $N=|\mathcal{Q}|$ be the number of qubits, $M=|\mathcal{S}|$ the number of storage sites, $P=|\mathcal{E}|$ the number of entanglement sites, $G$ the total number of two-qubit gates, and $m_\ell$ the number of qubit moves generated at stage $\ell$. In the current implementation, the routing-aware initial mapping consists of: (i) distance preprocessing over storage and entanglement sites, which costs $O(MP)$; (ii) priority-weight computation over the scheduled gates, which is linear in the schedule size; and (iii) assignment of qubits to storage sites using the AOD-conflict-aware score. The last step is dominant: for each qubit, the algorithm scans the remaining candidate storage sites and compares the induced movement vector against the already planned vectors. In the worst case, with $M=\Theta(N)$ and a linear number of previously planned vectors, this yields $O(N^3)$ time for the initial mapping phase.

The stage-wise placement logic is cheaper. Idle-qubit management scales with the number of qubits currently resident in the entanglement zone and the cost of querying the next two-qubit reuse, while active placement scales with the number of two-qubit gates in the stage and the search for a free entanglement pair. Routing then constructs dependency and conflict relations among the moves in each stage and applies a greedy MIS routine, giving a cost of $O\!\left(\sum_\ell m_\ell^2\right)$ in the worst case. Therefore, we do not claim that the full ZAP compiler is strictly linear in the most general asymptotic sense.

What our data support is a practical statement: ZAP is single-pass and non-iterative, so it avoids the repeated global search loops used by ZAC, Enola, and PowerMove. Moreover, the zoned architecture bounds the number of simultaneously relevant entanglement sites and limits the routing batch width in each stage. Under these hardware-imposed constraints, the effective values of $m_\ell$ remain small relative to the total circuit size, so the observed end-to-end compilation time grows close to linearly over the regime we evaluate. These curves do not constitute a formal asymptotic proof that no future intersection can occur outside the evaluated range, but they do indicate that ZAP has a substantially smaller effective slope and better practical scalability than the baselines over the problem sizes of interest.

\subsection{Ablation Studies}

\subsubsection{Strategy Ablation: Look-Ahead Trade-off}

\begin{figure}[t]
    \centering
    \includegraphics[width=0.99\columnwidth]{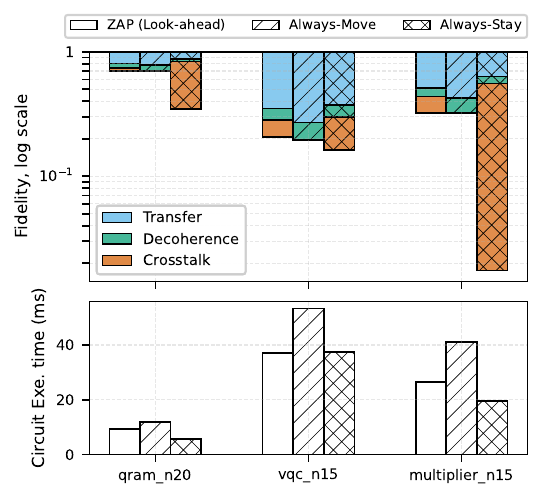}
    \caption{Ablation of the idle-qubit keep-versus-move policy, illustrating the crosstalk-versus-transport trade-off in ZAP's placement strategy. The top panel reports fidelity breakdowns on qram\_n20, vqc\_n15, and multiplier\_n15 for three variants: ZAP with the dynamic look-ahead rule, Always-Move, and Always-Stay. The bottom panel reports the corresponding circuit execution time. The dynamic policy captures most of the crosstalk suppression of Always-Move while avoiding much of its runtime overhead, and it avoids the severe crosstalk growth of Always-Stay on structured workloads.\label{fig:strategy_ablation}}
    \label{fig:strategy_ablation}
\end{figure}

The most informative ablation inside ZAP is the keep-versus-move decision for idle qubits in the entanglement zone. As described by the cost comparison above and Alg.~\ref{alg:stage_placement}, ZAP explicitly compares the predicted crosstalk penalty of leaving a qubit in place against the transfer and decoherence penalty of returning it to storage. To isolate the effect of this rule, we compare the full dynamic policy against two static baselines: \emph{Always-Move}, which aggressively returns every idle qubit to storage, and \emph{Always-Stay}, which keeps idle qubits in the entanglement zone whenever possible. Fig.~\ref{fig:strategy_ablation} shows that neither extreme is uniformly desirable. Always-Move suppresses crosstalk, but its larger number of transfers increases execution time and can reintroduce transfer and decoherence loss. Always-Stay minimizes movement, but it can accumulate substantial crosstalk on structured circuits with nonuniform reuse distances.

The dynamic look-ahead policy provides the best overall trade-off. On qram\_n20 and multiplier\_n15, ZAP attains fidelity close to Always-Move while noticeably reducing circuit execution time, showing that it captures most of the crosstalk benefit without paying the full shuttling cost. On multiplier\_n15, the gap relative to Always-Stay is especially dramatic: keeping idle qubits in place leads to a collapse in fidelity driven by crosstalk, whereas the dynamic policy selectively evacuates only the qubits for which the expected crosstalk cost exceeds the movement penalty. On vqc\_n15, where the three policies are closer, ZAP still achieves the strongest balance between the top-panel fidelity breakdown and the bottom-panel runtime. These results confirm that the benefit of look-ahead is not simply lower movement or lower crosstalk in isolation, but the ability to choose between them on a stage-by-stage basis.

\subsubsection{Hardware Parameter Sensitivity}

\begin{figure*}[t]
    \centering
    \includegraphics[width=1\linewidth]{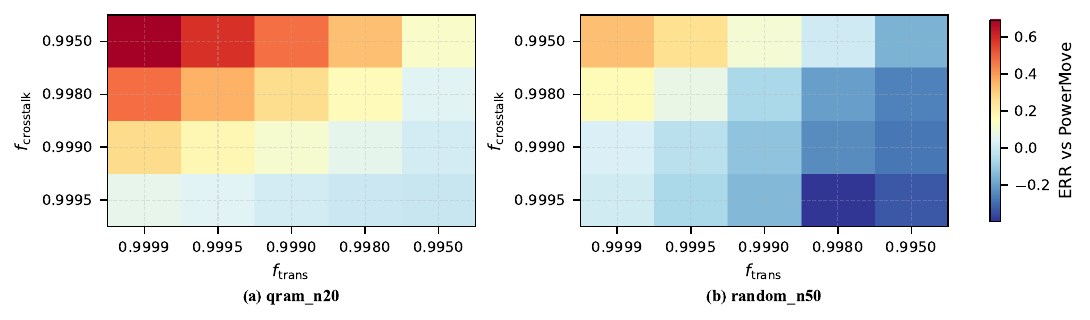}
    \caption{ERR of ZAP relative to PowerMove under different transfer and crosstalk fidelities. The heatmaps sweep $f_{\mathrm{tr}}$ and $f_{\mathrm{xtalk}}$ for (a) qram\_n20 and (b) random\_n50. Positive values indicate that ZAP achieves lower error than PowerMove. ZAP's advantage is strongest when crosstalk is costly and transfer is relatively reliable, and the effect is markedly stronger on the structured QRAM benchmark than on the topology-homogeneous random circuit.\label{fig:hardware_sensitivity}}
    \label{fig:hardware_sensitivity}
\end{figure*}
  
To test whether the benefit of the dynamic policy is robust beyond a single hand-tuned hardware setting, we sweep the transfer fidelity $f_{\mathrm{tr}}$ and the crosstalk fidelity $f_{\mathrm{xtalk}}$ and measure the resulting error reduction rate (ERR) of ZAP relative to PowerMove:
\begin{equation}
    \text{ERR} = \frac{E_\text{PowerMove} - E_\text{ZAP}}{E_\text{PowerMove}},
\end{equation}
where $E_\text{PowerMove}$ and $E_\text{ZAP}$ denote the error rates of PowerMove and ZAP, respectively, with error rate defined as $E=1-F$ for a given fidelity $F$.

We use qram\_n20 as a representative structured benchmark and random\_n50 as a topology-homogeneous random circuit. As shown in Fig.~\ref{fig:hardware_sensitivity}(a), the response on qram\_n20 is clear and interpretable rather than arbitrary: ZAP remains advantageous over a broad region of the parameter space, with the largest gains appearing when crosstalk is relatively severe and transport is relatively reliable. In that regime, the look-ahead policy can exploit the heterogeneous reuse structure of QRAM to suppress crosstalk without incurring unnecessary shuttling.

The random\_n50 heatmap is more moderate, as expected, and does not mirror the structured case. Random 3-regular circuits intentionally remove strong motifs, long-range reuse asymmetries, and other high-contrast features that make adaptive keep-versus-move decisions especially valuable. As a result, ZAP's relative advantage narrows and can become neutral or slightly negative in movement-dominated regimes where transfer loss is more important than crosstalk. We view this not as a contradiction but as a refinement of the claim: ZAP's dynamic policy is most valuable on structured workloads with heterogeneous reuse patterns, whereas on topology-homogeneous random circuits it remains competitive but has less structure to exploit. This interpretation is consistent with the earlier fidelity results, where the gaps among compilers are markedly larger on structured benchmarks than on random 3-regular circuits.

\section{Related work}\label{sec:related-work}
  
Compilation strategies must be tailored to specific quantum hardware due to differences in native gates and chip topologies~\cite{zhu2025quantum}. For superconducting platforms, methods such as the SABRE algorithm~\cite{li2019tackling}, randomized circuit compilation~\cite{hashim2020randomized}, and AI-based techniques~\cite{wang2024quantum} have been developed, and one-shot global optimization approaches such as ILP~\cite{wille2019mapping}, MaxSAT~\cite{molavi2022qubit}, and subgraph isomorphism~\cite{li2020qubit} have been used to solve placement and routing problems on superconducting platforms with fixed, local qubit connectivity. As ion-trap and neutral-atom systems increasingly adopt distributed and heterogeneous architectures to improve scalability and connectivity, recent frameworks have begun to address the associated compilation and communication challenges~\cite{xiang2025adaptdqc,tao2025qtenon}. Resource-virtualized and hardware-aware compilation systems, such as QSteed, further improve real-device compilation and task management for practical quantum computing backends~\cite{xu2025qsteed}. Strategies leveraging ion shuttling---such as TILT for linear traps~\cite{wu2021tilt,wu2024boss} and various methods for QCCD architectures~\cite{stevens2017automating,wu2019ilp, zhu2025s}---primarily focus on efficient ion transport and mitigating performance drops as qubit numbers increase. 

For neutral atom platforms, parallel qubit movement and global laser operations enable high connectivity and parallelism, which have inspired tailored compilation schemes. Early works such as OLSQ-DPQA~\cite{tan2024compiling} focus on optimal layout synthesis using SMT solvers. Enola~\cite{tan2024compilation} further improves scheduling through graph edge coloring, reducing stages by 3.7$\times$ and improving fidelity by up to 5.9$\times$ over prior methods. Other works~\cite{baker2021exploiting} study atom loss errors in flexible atom arrays with long-distance interactions. A representative line of prior work focuses on non-zoned neutral-atom architectures with flexible qubit arrays and global addressing constraints, including Geyser~\cite{patel2022geyser}, FPQA-C~\cite{wang2023fpqa}, Q-Pilot~\cite{wang2023q}, Atomique~\cite{wangAtomiqueQuantumCompiler2024}, Qompose~\cite{silver2024qompose}, and Enola~\cite{tan2024compilation}. In parallel, a distinct line of later work studies zoned neutral-atom architectures that explicitly separate storage and entanglement regions, including NALAC~\cite{stade2024abstract}, Arctic~\cite{decker2024arctic}, ZAC~\cite{lin2025reuse}, and PowerMove~\cite{ruan2024powermove}. These zoned approaches improve scalability and mitigate crosstalk through structured atom movement, scheduling, and reuse management. More recent follow-up works further integrate routing-aware and AOD-aware placement into zoned compilation, including the routing-aware placement method of Stade \emph{et al.}~\cite{Stade_2025} and the later scalable search framework of Stade \emph{et al.}~\cite{stade2025searchsmarterharderscalable}.

Collectively, these frameworks address the compilation and scheduling challenges of reconfigurable neutral atom architectures, enhancing scalability, fidelity, and efficiency. Our work primarily compares with Enola~\cite{tan2024compilation}, which requires gate commutativity, ZAC~\cite{lin2025reuse}, a reuse-aware compiler with iterative placement loops, and PowerMove~\cite{ruan2024powermove}, a strong recent compiler that utilizes a multi-stage heuristic framework to co-optimize gate scheduling and continuous atom routing. In contrast to all three approaches, our partitioned architecture and deterministic single-pass compiler minimize parallel atomic movement, thereby improving fidelity and substantially reducing compilation time without iterative overhead, offering a more robust and scalable approach to neutral atom quantum computing.
  
\section{Discussion and Conclusion}\label{sec:conclusion}
  
In this work, we presented ZAP, a co-designed zoned architecture and deterministic compiler for neutral-atom quantum computing. ZAP combines a hardware-structured storage--entanglement layout with ASAP-separate scheduling, routing-aware placement, cost-aware idle-qubit management, and conflict-aware routing. Rather than relying on repeated global search, the compiler resolves scheduling, placement, and transport in a single-pass flow that is explicitly matched to the constraints of AOD-based atom movement. The main contribution is therefore not an isolated heuristic, but the overall co-design between zoned hardware structure and a non-iterative compilation strategy.

Our evaluation on structured benchmarks and random 3-regular circuits shows that this co-design yields substantial practical benefits. Relative to the zoned baselines ZAC and PowerMove, ZAP typically reduces compilation time from tens of seconds to below 0.1~s and delivers speedups exceeding 1,000$\times$, while remaining competitive or superior in execution quality. Relative to the non-zoned Enola baseline, the compilation speedup exceeds 10,000$\times$ on the evaluated suite, while fidelity improves markedly because zoning strongly suppresses crosstalk. Among zoned compilers, ZAP's fidelity advantage is most pronounced on structured workloads with irregular connectivity and nonuniform qubit reuse, where better scheduling, placement, and transport decisions accumulate over many stages; on random circuits, the fidelity gap is smaller but ZAP remains competitive and preserves the same scalability advantage. Importantly, these fidelity improvements do not come at the cost of longer physical runtime: across zoned baselines, ZAP is consistently competitive in execution time and is often the fastest because it reduces unnecessary shuttling and waiting overhead.

We also emphasize a practical systems perspective. For a one-off NISQ experiment with only tens to a few hundred qubits, reducing classical compilation time from minutes to milliseconds is not always the dominant end-to-end bottleneck compared with quantum runtime, hardware availability, and experimental overhead. In standard workflows, compilation is typically performed before atom loading and execution, so its main benefit is often improved experimental throughput rather than directly extending the coherence budget of a single shot. Even so, fast compilation is highly valuable in iterative and closed-loop settings such as repeated benchmarking, calibration-aware recompilation, parameter sweeps, and adaptive workflows in which many related circuit instances must be regenerated under changing hardware conditions. In such scenarios, search-based latency can accumulate into a practical control-loop bottleneck. This systems consideration becomes even stronger in fault-tolerant settings, where the compiled objects shift from physical qubits to logical qubits or logical patches and repeated scheduling, routing, and resource-allocation decisions can themselves become a throughput limiter. It is also important from the perspective of quantum cloud services: because compilation and transpilation lie on the user-facing path before job submission, long compile latency degrades interactive feedback and makes the service less friendly to online users. This practical concern is reflected in existing cloud ecosystems such as IBM Qiskit and the Quafu platform, which already expose transpilation or compilation as part of the cloud workflow~\cite{javadiabhari2024quantumcomputingqiskit,quafu2024docs}. Although neutral-atom hardware has not yet been widely deployed as a public cloud service, the same requirement for low user-visible compilation latency will become increasingly important once such platforms are opened to online access.

More broadly, ZAP supports the view that hardware-structured, non-iterative compilation is a promising path toward scalable and noise-aware neutral-atom computing. As hardware coherence and array size continue to improve, the relative importance of compiler decisions about crosstalk, transport, and waiting time will only increase. Future work includes exploring the co-design space of storage and entanglement zones, further reducing atom-transfer overhead, and extending the same philosophy toward fault-tolerant compilation. In that longer-term setting, the compilation objects would need to be lifted from physical atoms to logical qubits or logical patches, and the cost model would need to shift from physical movement and decoherence to logical communication, distillation resources, and code-cycle latency. Although such an extension remains future work, the central principle of replacing repeated global search with hardware-aware deterministic compilation remains applicable.

\section*{Acknowledgments}

We sincerely thank Chenghong Zhu, Xudan Chai, and Yanwu Gu for their invaluable discussions. This work was supported by the National Key Research and Development Program of China (Grant No. 2025YFE0217200), the Beijing Natural Science Foundation (Grant No. Z220002), the National Natural Science Foundation of China (Grant No. 92365111), the Innovation Program for Quantum Science and Technology (Grant No. 2021ZD0302400), and the National Natural Science Foundation of China (Grant No. 92365206). ZAP is open source at~\url{https://github.com/BAQIS-Quantum/neutral-atom-compilation}. All data and code used in this work are publicly available in the same repository.

\bibliographystyle{ieeetran}
\bibliography{refs}

\end{document}